\documentclass[sn-mathphys-num]{sn-jnl}

\usepackage{graphicx}%
\usepackage{amsmath,amssymb,amsfonts}%
\usepackage{multirow}%
\usepackage{makecell}
\usepackage{booktabs}%
\usepackage{dsfont}
\usepackage{xcolor}%
\usepackage{textcomp}%
\usepackage{lmodern}

\begin{document}

\title[DyneTrion]{DyneTrion: A Spatio-temporally Coherent Generative Emulator for Protein Dynamics Across Timescales}

\author[1,2]{\fnm{Kaihui} \sur{Cheng}}
\equalcont{These authors contributed equally to this work.}

\author[2]{\fnm{Zhiqiang} \sur{Cai}}
\equalcont{These authors contributed equally to this work.}

\author[2]{\fnm{Peng} \sur{Tu}}

\author[2]{\fnm{Yisong} \sur{Yao}}

\author[1,2]{\fnm{Limei} \sur{Han}}

\author[2,3,4,5]{\fnm{Libo} \sur{Wu}}

\author*[1,2]{\fnm{Siyu} \sur{Zhu}}\email{siyuzhu@fudan.edu.cn}

\author*[2]{\fnm{Tzuhsiung} \sur{Yang}}\email{yangzixiong@sais.org.cn}

\author*[1,2]{\fnm{Yuan} \sur{Qi}}\email{qiyuan@fudan.edu.cn}

\affil[1]{\orgname{Artificial Intelligence Innovation and Incubation Institute, Fudan University}, \orgaddress{\city{Shanghai}, \country{China}}}

\affil[2]{\orgname{Shanghai Academy of AI for Science}, \orgaddress{\city{Shanghai}, \country{China}}}

\affil[3]{\orgname{School of Data Science, Fudan University}, \orgaddress{\city{Shanghai}, \country{China}}}

\affil[4]{\orgname{Institute for Big Data, Fudan University}, \orgaddress{\city{Shanghai}, \country{China}}}

\affil[5]{\orgname{MOE Laboratory for National Development and Intelligent Governance, Fudan University}, \orgaddress{\city{Shanghai}, \country{China}}}


\abstract{

Proteins function through coordinated motion across multiple spatial and temporal scales, underpinning processes such as ligand binding, allostery, and catalysis. However, accessing long-timescale conformational change through molecular dynamics (MD) simulations remains prohibitively expensive for systematic exploration across diverse systems. Here, we present DyneTrion, a generative protein dynamics emulator that jointly enforces geometric symmetry, structural consistency and temporal coherence within a single framework. DyneTrion uses a tri-attention architecture that integrates invariant point attention (IPA) for SE(3)-robust geometric updates, spatial attention anchored to a reference conformation to preserve structural integrity, and temporal attention to model correlated evolution across time frames.

Across 100-ns MD trajectory simulation benchmarks, DyneTrion reproduces MD-derived flexibility, ensemble distributions and interaction observables while maintaining stereochemical validity during extrapolation. To evaluate long time-scale generalization, we introduce dynamicPDB, a dataset of over 10,000 proteins with up to 1-µs all-atom trajectories at 10-ps resolution and accompanying physical annotations. On microsecond trajectories, DyneTrion preserves free-energy landscapes and metastable-state populations, and it supports large conformational propagation in apo-to-holo transitions and fast folders. Together, DyneTrion provides a scalable path from static structure prediction toward time-resolved, ensemble-faithful protein modeling. The code is publicly available at \url{https://github.com/fudan-generative-vision/DyneTrion}.
}


\maketitle
\section{Introduction}\label{sec1}
Protein function is intimately linked to three-dimensional structure, as structural organization underlies mechanisms such as molecular recognition, catalysis, and allosteric regulation. Experimental techniques including X-ray crystallography have enabled the determination of over 200,000 protein structures, providing a foundational view of protein architecture~\cite{burley2019rcsb}. 

More recently, advances in deep learning have transformed structural biology. Methods such as the AlphaFold series~\cite{jumper2021highly,abramson2024accurate}, the ESM family~\cite{rives2021biological,zeming23llm,hayes2025simulating}, and RoseTTAFold~\cite{baek21rose} have demonstrated that static protein structures can be predicted at near-experimental accuracy directly from amino acid sequences, expanding structural coverage to hundreds of millions of proteins.

While predicting static structures is instrumental, proteins are inherently dynamic entities whose biological function arises from motion across rugged conformational landscapes. These motions span fast, local fluctuations that modulate flexibility and interaction propensities~\cite{karplus1990molecular,teilum2009functional}, as well as slow, rare, barrier-crossing transitions that drive large-scale conformational rearrangements, reorganize interaction networks, and transiently expose or occlude functional sites~\cite{klepeis2009long,bahar2010global}.
Capturing such dynamics is essential for understanding catalysis, signaling, folding, and ligand recognition. Although all-aton molecular dynamics (MD) simulations provide a principled physical framework for modeling protein motion, their computational cost severely limits systematic exploration across proteins, timescales, and conditions—particularly when long-timescale processes and rare events dominate functional behavior~\cite{hospital2015molecular,hollingsworth2018molecular}. Additionally, existing MD datasets suffer from limited availability, diversity, and heterogeneity; for example, Atlas~\cite{vander23atlas} provides thousands of dynamic proteins with a 10 picosecond sampling interval, but the dataset is limited in functional and structural diversity. MoDEL~\cite{meyer10model} and Dynameomics~\cite{kehl08dynameomics} have primarily offered limited-scale MD datasets for soluble proteins. Other datasets, such as Medusa~\cite{vander2021medusa} and Vander et al.~\cite{marchetti2021machine} are tailored to specific tasks like flexibility prediction and contain only a few hundred proteins.

Motivated by these limitations, recent deep-learning approaches have emerged as generative surrogates for MD simulations of biomolecular motions. Score-based models have been adapted to emulate short-time protein dynamics by learning conditional distributions over atomic coordinates: MDGEN~\cite{jing2024generative} models short-horizon all-atom protein dynamics using diffusion-based generation of trajectory segments conditioned on an initial conformation, while UniSim~\cite{yu2025unisim} emulates protein motion by learning equivariant stochastic transitions that propagate structures forward over coarse-grained time steps. In parallel, autoregressive approaches propagate protein motion in discretized structure space; for example, ProTDyn~\cite{liu2025protdyn} defines a factored autoregressive likelihood over residue-level structure tokens across time, jointly capturing Boltzmann-consistent equilibrium ensembles and multi-timescale transitions. Complementarily, ensemble-based methods focus on sampling structural conformations without explicit temporal modeling: AlphaFlow~\cite{jing2023alphaflow} reformulates static structure predictors such as AlphaFold~\cite{jumper2021highly} and ESMFold~\cite{rives2021biological} as flow-matching conformation generators, Str2Str~\cite{lu2024strstr} enables zero-shot structure-to-structure translation, and BioEmu~\cite{bioemu2025} employs conditional denoising diffusion over AlphaFold-derived representations to directly sample Boltzmann-consistent equilibrium ensembles from large-scale MD and experimental data.

Despite notable progress, existing approaches typically prioritize only a subset of the requirements for faithful trajectory generation—geometric equivariance, structural consistency, or temporal coherence—rather than addressing all three simultaneously. A practical dynamics emulator needs to respect local rotational and translational symmetries to prevent rigid-body drift, preserve long-range packing and topology as conformations evolve, and model smooth, correlated motion over time to support long time-scale extrapolation. When any one of these elements is missing, characteristic failure modes emerge: score-based generators without explicit temporal modeling can drift or decorrelate across time frames; locally equivariant updates without a global spatial anchor can accumulate long-range packing errors; and generations that do not explicitly enforce local symmetry or geometry tend to degrade stereochemical validity over long time-scale. Accordingly, score-based trajectory models often accrue geometric artifacts during extrapolation, autoregressive models may attenuate realistic fluctuation amplitudes or rare transitions, and ensemble-only methods lack a principled notion of dynamical continuity—leading to increasingly degraded performance as the prediction horizon extends, especially when trajectories must traverse free-energy barriers or capture slow collective motions.

To address this gap, we introduce DyneTrion, a protein dynamics emulator built on a tri-attention architecture that explicitly addresses these challenges. DyneTrion integrates (i) an invariant point attention (IPA) mechanism to preserve SE(3)-invariant aggregation while referencing local frames, preventing rigid-body drift and stabilizing local geometry, (ii) a spatial attention module to maintain global structural consistency relative to the reference conformation, and (iii) a temporal attention module to model correlated motion across time frames and enforce temporal coherence. By jointly attending over structural and temporal information, DyneTrion learns MD-consistent dynamical statistics rather than memorizing individual trajectories, enabling accurate short-timescale emulation and stable long-timescale extrapolation within a unified generative framework.

We show that DyneTrion accurately reproduces MD-derived dynamical observables on nanosecond timescales, maintains stereochemical validity during long-horizon extrapolation, and recovers free-energy landscapes and metastable-state populations on microsecond scales. Beyond trajectory generation, DyneTrion also functions as a general ensemble generator, achieving performance comparable to specialized ensemble-sampling models. To evaluate long-time generalization, we  introduce dynamicPDB, a large-scale dataset of over 10,000 proteins with up to 1-µs all-atom MD trajectories at 10-ps resolution and rich physical annotations of temperature, force, velocities, and energies, providing a principled benchmark for long-horizon extrapolation and dynamical fidelity. Together, DyneTrion and dynamicPDB bridge the gap between static structure prediction and time-resolved molecular behavior.

\section{Results}

\subsection{Dynamic emulator with a Tri-attention Architecture.}

\paragraph{Model Architecture}

\begin{figure}[!t]
  \centering
  \includegraphics[width=1.0\linewidth]{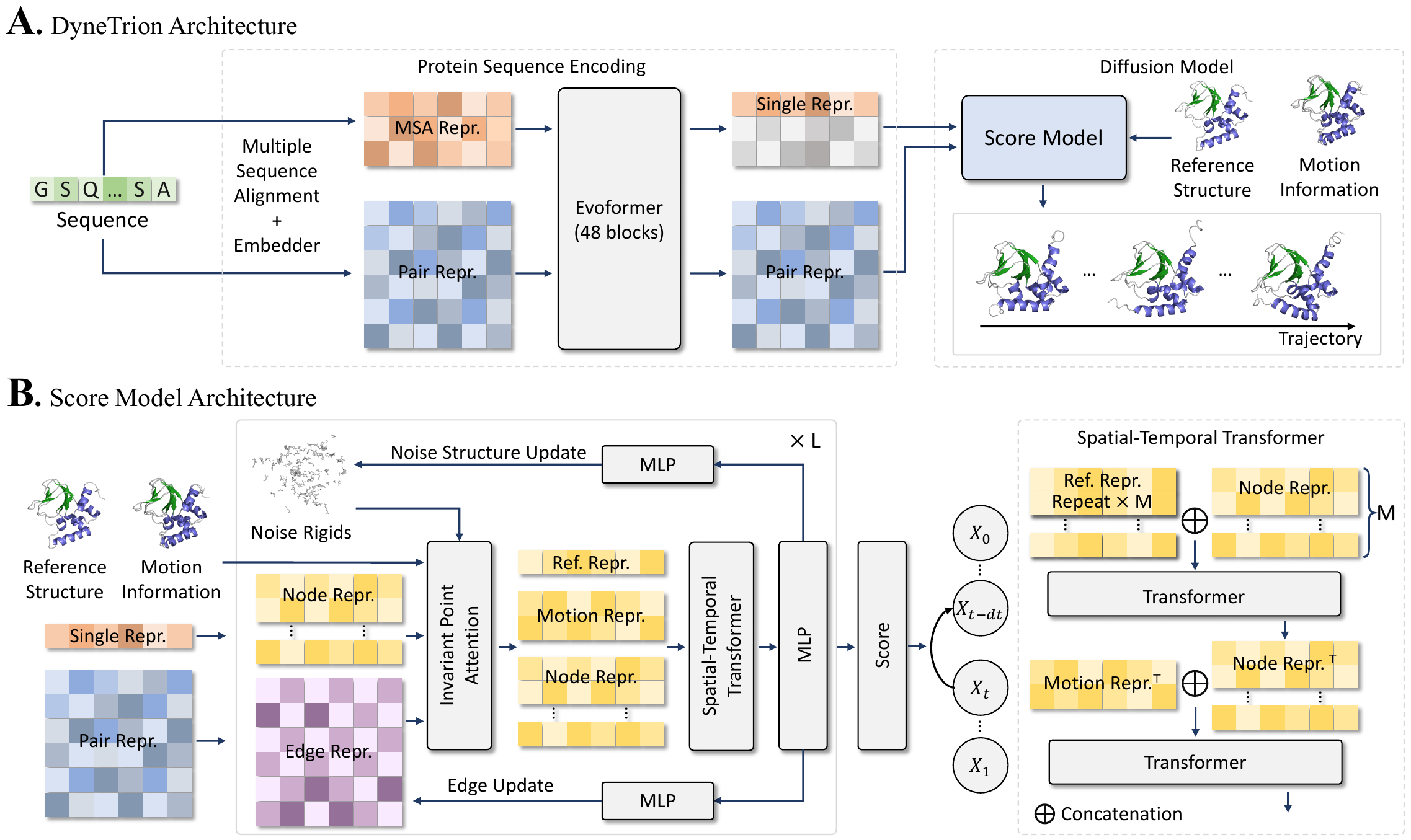} %
  \caption{ \textbf{Overview of architectures.} (A) The architecture of DyneTrion. Given an input sequence, DyneTrion derives representations through multiple sequence alignment and the Evoformer, then employs a score model conditioned on a reference structure and motion information to generates a protein dynamic trajectory. (B) The architecture of the score model, which comprises invariant point attention, spatial-temporal transformers, and multiple linear projection. Protein trajectories are represented using rigid-body transformations.
  }
  \label{fig:pipeline}
\end{figure}

We developed DyneTrion, a score-based generative model for protein Dynamics emulation using a Tri-attention architecture. DyneTrion uses the AlphaFold2 Evoformer\cite{jumper2021highly} to encode the input sequence into single and pair representations. These representations serve as input to a diffusion model together with a reference structure and motion information carried by two consecutive frames (Fig~\ref{fig:pipeline} A). The denoising diffusion model is employed to estimate scores~\cite{song2019generative} and iteratively denoise the input noise, generating protein dynamic trajectories (Fig~\ref{fig:pipeline} B).  

Within each diffusion step, node and edge representations are processed by the IPA module conditioned on Gaussian-initialized noisy rigid-body frames, reference structures and motion information, producing structure-aware feature updates. The updated representations are further integrated by the ST, TT modules and an MLP to predict incremental updates to the noisy rigids and edge representations. 
This process is repeated for $L$ times to estimate the score function at each denoising step, and eventually generate coherent protein dynamic trajectories (Fig.~\ref{fig:pipeline}B). 

\paragraph{Tri-Attention Enables Physically Consistent Protein Dynamics}

\begin{figure}[t!]
  \centering
  \includegraphics[width=1.0\linewidth]{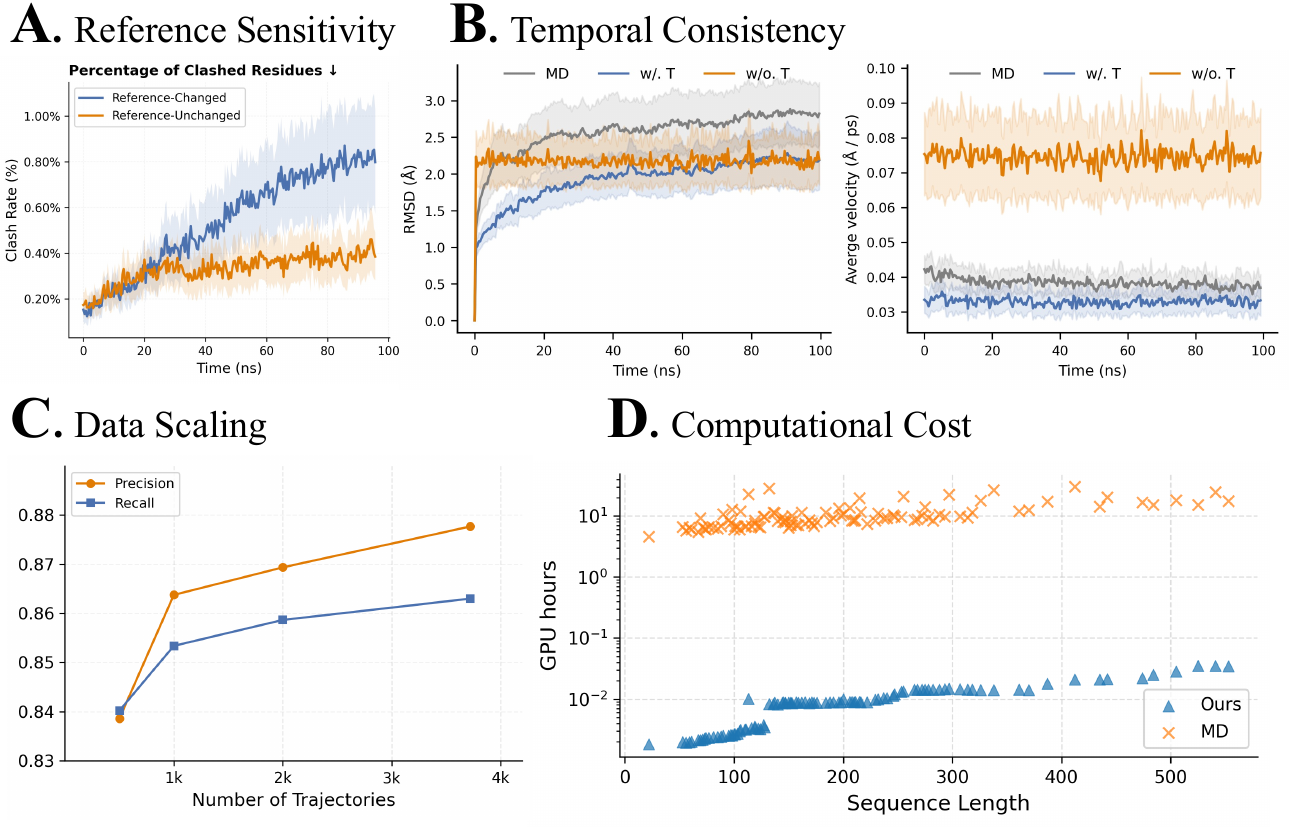} %
  \caption{\textbf{Ablation and scalability analysis of DyneTrion.}
  (A) Reference sensitivity analysis. Percentage of clashed residues over time as a function of variations in the reference structures.
(B) Temporal consistency analysis. Left: C$\alpha$–RMSD relative to the initial structure for MD, the full model with temporal transformer (w/ T), and an ablated variant without temporal modeling (w/o T). Right: Average velocity between consecutive frames, reporting mean and variance across trajectories.
(C) Data-scaling behavior of DyneTrion, showing structure coverage precision and recall as a function of the number of training trajectories.
(D) GPU hour analysis. Runtime comparison between Ours and MD(normalized to 100-ns simulation).
} 
  \label{fig:ablation}
\end{figure}

The spatial–temporal transformer in DyneTrion is designed to jointly encode spatial context from the reference structure and temporal dependencies from protein motion. DyneTrion relies on a dedicated spatial module to preserve stereochemical validity throughout long-horizon generation. As shown in Fig.~\ref{fig:ablation}A, conditioning the model on a fixed reference structure substantially reduces the fraction of steric clashes over time. When the reference structure is changed iteratively as trajectory generation propagates, the percentage of clashed residues increases steadily, whereas retaining a high-quality initial structure (e.g., an experimentally resolved PDB conformation) maintains stereochemical consistency across the trajectory. 

To assess the role of the temporal module, we performed an ablation study in which the temporal transformer was removed. As shown in Fig.~\ref{fig:ablation}B, the full model exhibits a smooth and stable increase in RMSD relative to the first frame, closely matching the gradual divergence observed in MD trajectories. In contrast, removing the temporal transformer leads to a rapid RMSD increase at early time steps followed by pronounced fluctuations, indicating a loss of temporal coherence. Consistent trends are observed in per-frame average velocities and their variances, which remain within the MD-derived range only when the temporal module is present.

We further evaluated the data-scaling behavior of DyneTrion by varying the number of training trajectories while keeping the same set of 100 test proteins. As shown in Fig.~\ref{fig:ablation}C, both precision and recall for structure coverage improve monotonically with increased training data, indicating that DyneTrion benefits from additional dynamical examples rather than saturating or overfitting. 

Finally, we compare the computational cost of DyneTrion against classical MD simulations. As shown in Fig.~\ref{fig:ablation}D, DyneTrion achieves a two to three orders-of-magnitude reduction in GPU time across a wide range of protein sequence lengths on 1 GPU. This dramatic speedup enables rapid emulation of protein dynamics across multiple temporal scales—from short-time local fluctuations to long-time conformational rearrangements—making high-throughput and long-horizon dynamical analysis feasible on commodity GPU hardware.

\subsection{Short-Timescale Trajectory Emulation.}

\paragraph{Benchmarking against SOTA methods}

We benchmark DyneTrion against state-of-the-art (SOTA) methods by evaluating both dynamical properties and structural quality of the generated trajectories relative to 100-ns MD trajectories. Following the metrics proposed by Jing, \textit{et al.}~\cite{jing2023alphaflow} and Lu, \textit{et al.}~\cite{lu2024strstr}, we assess short-timescale dynamics using metrics that probe five complementary aspects of protein motion.

To evaluate protein flexibility, we measure pairwise C$\alpha$-RMSD and residue-level RMSF, testing whether DyneTrion reproduces both global conformational variability and spatially resolved fluctuations observed in MD. Distributional accuracy is assessed using root mean Wasserstein distance (RMWD), PCA-based Wasserstein distances and cosine similarity (MD PCA, Joint PCA, and PC-sim), which quantify how well the generated ensemble matches the MD distribution and recovers dominant collective motions beyond framewise errors. Ensemble observables, including weak/transient contacts and solvent exposure, probe whether the model captures residue-level interaction patterns and intermittent structural rearrangements characteristic of realistic protein dynamics.

In addition, structural quality is evaluated as following: precision and recall based on local distance difference test using $\mathrm{C}\alpha$ atoms (lDDT$_{\mathrm{C}\alpha}$) quantify ensemble coverage of MD conformations, balancing accuracy and diversity, while backbone distance constraints and steric-clash statistics measure stereochemical validity. The results are shown in Table~\ref{tab:performance}.

Quantitatively, DyneTrion attains the highest overall agreement with MD across all evaluated metrics. It achieves the highest correlations for flexibility measures, including pairwise RMSD ($r=0.831$), global RMSF ($r=0.803$), and per-target RMSF ($r=0.791$), indicating accurate recovery of both global and residue-resolved fluctuations. For distributional accuracy, DyneTrion consistently yields lower Wasserstein distances than other methods, with RMWD reduced to 1.950 compared to 2.768 for MDGEN and larger discrepancies for UniSim and ProTDyn, reflecting closer alignment of generated and MD ensembles. Improvements are similarly observed in PCA-based metrics, including MD PCA $\mathcal{W}_2$ (0.962), joint PCA $\mathcal{W}_2$ (1.480), and principal-component similarity (0.281), indicating more faithful recovery of dominant collective motions.

For ensemble observables, DyneTrion achieves higher Jaccard similarity (J) for weak contacts (0.504) and transient contacts (0.434), suggesting improved modeling of both persistent and dynamically formed residue interactions. Agreement on solvent exposure is likewise higher (exposed residue $J=0.528$), indicating more faithful reproduction of side-chain burial–exposure transitions. In terms of structural quality, DyneTrion attains the highest precision (0.878) and recall (0.863) for structure coverage, while maintaining near-ideal stereochemical validity across backbone distances and steric clash metrics, comparable to or exceeding competing approaches.

\begin{table}[!t]

\centering

\resizebox{\textwidth}{!}{%
\begin{tabular}{lccccccccccccccc}
\toprule
\multirow{4}{*}{\textbf{Model}} 
& \multicolumn{10}{c}{\textbf{Dynamic Properties}}
& \multicolumn{5}{c}{\textbf{Structural Quality}} \\
\cmidrule(lr){2-11} \cmidrule(lr){12-16}

& \multicolumn{3}{c}{\textbf{Predicting Flexibility}}
& \multicolumn{4}{c}{\textbf{Distributional Accuracy}}
& \multicolumn{3}{c}{\textbf{Ensemble Observables}}
& \multicolumn{2}{c}{\textbf{Structure Coverage}}
& \multicolumn{3}{c}{\textbf{Stereochemical Validity}} \\
\cmidrule(lr){2-4}
\cmidrule(lr){5-8}
\cmidrule(lr){9-11}
\cmidrule(lr){12-13}
\cmidrule(lr){14-16}

& \makecell{Pairwise \\ RMSD $r\uparrow$}
& \makecell{Global \\ RMSF $r\uparrow$}
& \makecell{Per-target \\ RMSF $r\uparrow$}
& \makecell{RMWD \\ $\downarrow$}
& \makecell{MD PCA \\ $\mathcal{W}_2\downarrow$}
& \makecell{Joint PCA \\ $\mathcal{W}_2\downarrow$}
& \makecell{PC-sim \\ $\uparrow$}
& \makecell{Weak \\ contacts $J\uparrow$}
& \makecell{Transient \\ contacts $J\uparrow$}
& \makecell{Exposed \\ residue $J\uparrow$}
& \makecell{Precision \\ $\uparrow$}
& \makecell{Recall \\ $\uparrow$}
& \makecell{Val-C$_\alpha$-C$_\alpha$ \\ $\uparrow$}
& \makecell{Val-C-N \\ $\uparrow$}
& \makecell{Val-Clash \\ $\uparrow$} \\
\midrule

\addlinespace[2pt]
\multicolumn{16}{c}{\textit{Score-based models}} \\
\addlinespace[4pt]

\textbf{DyneTrion}
& \textbf{0.831} & \textbf{0.803} & \textbf{0.791}
& \textbf{1.950} & \textbf{0.962} & \textbf{1.480} & \textbf{0.281}
& \textbf{0.504} & \textbf{0.434} & \textbf{0.528}
& \textbf{0.878} & \textbf{0.863}
& \textbf{99.952} & \textbf{99.898} & 99.675 \\

\textbf{MDGEN}
& 0.700 & 0.662 & 0.754
& 2.768 & 1.015 & 2.424 & 0.147
& 0.415 & 0.294 & 0.442
& 0.743 & 0.855
& 98.990 & 98.844 & 99.492 \\

\textbf{UniSim}\textsuperscript{a}
& 0.292 & 0.318 & 0.469
& 4.296 & 1.243 & 3.915 & 0.103
& 0.286 & 0.099 & 0.333
& 0.649 & 0.849
& 99.153 & 98.759 & \textbf{99.983} \\

\addlinespace[6pt]
\multicolumn{16}{c}{\textit{Autoregressive models}} \\
\addlinespace[4pt]

\textbf{ProTDyn}\textsuperscript{b}
& 0.122 & 0.136 & 0.508
& 10.069 & 2.289 & 8.423 & 0.117
& 0.262 & 0.127 & \text{N/A}\textsuperscript{c}
& 0.518 & 0.851
& 99.419 & 99.266 & 99.940 \\
\bottomrule
\end{tabular}%
}
\caption{\textbf{Benchmarking DyneTrion against state-of-the-art models on protein dynamics emulation.} Performance is evaluated on 100 test proteins using metrics spanning dynamic properties and structural quality. DyneTrion achieves the highest overall agreement with MD across flexibility, distributional accuracy, and ensemble observables, while maintaining high structure coverage and stereochemical validity. Arrows indicate whether higher ($\uparrow$) or lower ($\downarrow$) values correspond to better performance. Detailed definitions and computation procedures are provided in Section~\ref{sec:metric}.}
\label{tab:performance}
\vspace{2pt}
\begin{minipage}{\textwidth}\footnotesize
\textsuperscript{a} UniSim results are reported for the energy-minimization model (UniSim-em). \\
\textsuperscript{b} ProTDyn is evaluated using the backbone-only model with pretrained weights. \\
\textsuperscript{c} Exposed-residue $J$ is not available for ProTDyn.
\end{minipage}
\end{table}

\begin{figure}[t!]
  \centering
  \includegraphics[width=1.0\linewidth]{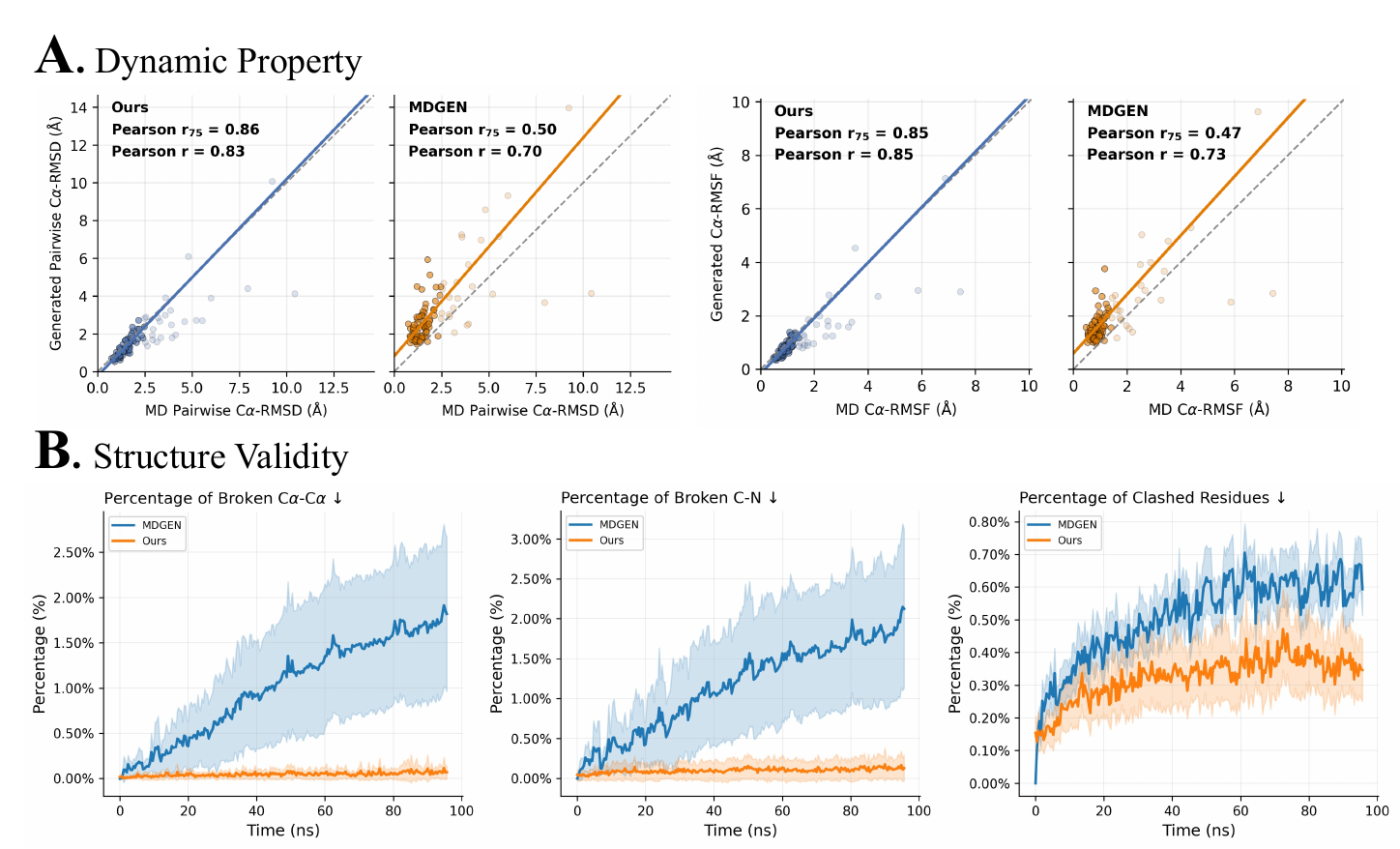} %
  \caption{ \textbf{DyneTrion captures short-timescale protein dynamics with high fidelity and temporal consistency.}
(A) Pointwise comparison of MD-derived and model-generated pairwise C$\alpha$–RMSD and C$\alpha$–RMSF, where pairwise RMSD is defined as the average RMSD between pairs of frames within a trajectory.
(B) Structural validity as a function of generated trajectory length, quantified by the percentage of broken backbone distances (excessively long inter-residue C$\alpha$–C$\alpha$ and C–N distances) and the percentage of clashed residues.
} 
  \label{fig:MD_emulation}
\end{figure}

\paragraph{Diagnostic Comparison with the Strongest Baseline}
Having established DyneTrion’s superior aggregate performance across all evaluated metrics (Table~\ref{tab:performance}), we next perform a more fine-grained comparison against MDGEN, which consistently represents the highest baseline among existing SOTA methods. This analysis examines model behavior at the level of individual proteins, enabling us to distinguish systematic biases from random errors and to assess robustness across different dynamical regimes.

Figure~\ref{fig:MD_emulation}A presents per-protein comparisons of pairwise C$\alpha$–RMSD and C$\alpha$–RMSF against MD, enabling a regime-dependent analysis of model behavior. For proteins whose MD-derived RMSD and RMSF fall within the lower to intermediate range of the test distribution (approximately the first three quartiles), DyneTrion exhibits consistently higher Pearson correlations and closely follows the MD trend ($\mathrm{r}_{75}$=0.86 v.s. 0.50). In this regime, MDGEN displays a systematic bias, tending to overestimate dynamical amplitudes in both metrics.

For proteins characterized by larger RMSD and RMSF values, corresponding to more flexible or dynamically heterogeneous systems, DyneTrion produces systematically more conservative predictions, avoiding the large excursions observed in MDGEN. Notably, MDGEN does not exhibit a clear or consistent trend in this high-dynamics regime, leading to increased scatter and reduced correlation with MD, as evidenced by the big difference between its $\mathrm{r}_{75}$ and $\mathrm{r}$ values. Together, these results indicate that DyneTrion maintains accurate scaling of dynamical amplitudes within the dominant regime while favoring stability and physical plausibility when extrapolating to highly flexible proteins in the tail of the test distribution.

\paragraph{Stereochemical Stability under Extrapolation}
Beyond time-averaged dynamical accuracy, we evaluate how well the generated trajectories preserve stereochemical consistency as generation extrapolates in time (Fig.~\ref{fig:MD_emulation}B). Local backbone integrity is assessed by the percentage of unphysically long neighboring-residue C$\alpha$–C$\alpha$ and C–N distances, which indicate backbone breakage. Across all extrapolation horizons, DyneTrion produces substantially fewer such violations than MDGEN, demonstrating improved preservation of local structural geometry. Moreover, DyneTrion exhibits markedly smaller variance across test proteins, indicating greater robustness with respect to protein identity.

We further assess spatial consistency using the percentage of clashed residues, a metric sensitive to both local packing and long-range residue positioning. DyneTrion consistently yields lower clash rates than MDGEN, indicating improved global structural organization. Notably, as extrapolation time increases, the clash rate for DyneTrion plateaus, whereas MDGEN continues to accumulate clashes, suggesting progressive structural degradation. Together, these results show that DyneTrion maintains stereochemical stability over long extrapolation times, preserving both local backbone integrity and global residue packing more effectively than MDGEN.

\subsection{Long-Timescale Extrapolation.}

Emulating protein dynamics over microsecond timescales poses qualitatively different challenges from short-timescale prediction. At these horizons, trajectories must cross free-energy barriers, sample rare metastable states, and preserve slow collective motions that define the conformational ensemble. The availability of microsecond-scale trajectories in dynamicPDB enables a systematic evaluation of whether generative models can faithfully extrapolate beyond nanosecond regimes, capturing both ensemble-averaged thermodynamics and long-time dynamical properties.

\begin{figure}[!tp]
  \centering
  \includegraphics[width=1.0\linewidth]{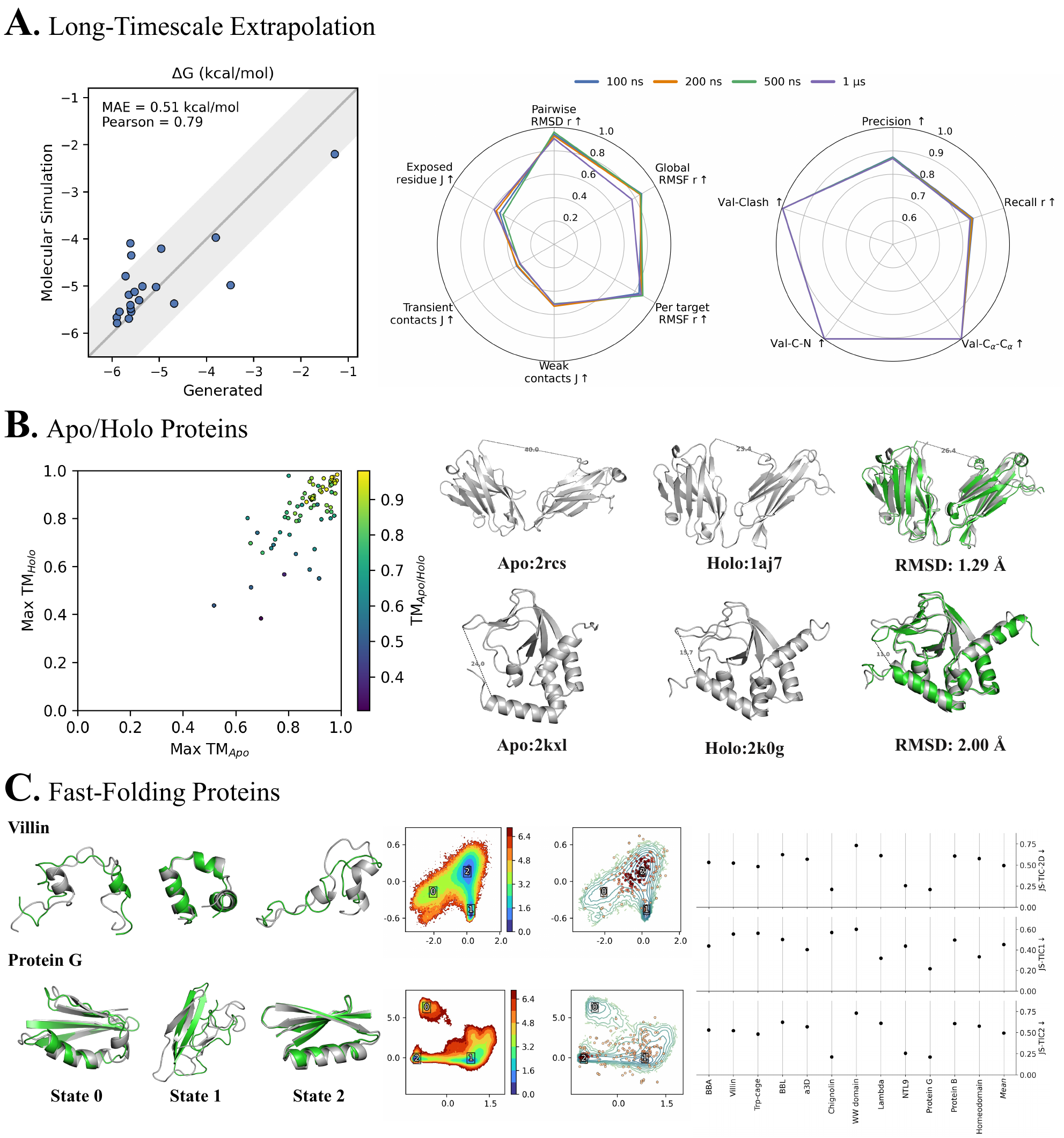} %
  \caption{\textbf{DyneTrion enables long-timescale protein dynamics extrapolation.}
(A) Ensemble-level agreement between MD and model-generated free energies ($\Delta G$) on microsecond trajectories from dynamicPDB, with MAE and Pearson/Spearman correlations reported. Right: stability of dynamical observables across extrapolation lengths (100 ns–1 $\mu$s), including RMSD, RMSF, and contact-based metrics.
(B) Apo-to-holo transitions. Left: maximin TM-score between generated structures and experimental apo and holo conformations (apo–holo TM-score encoded by color). Right: representative generated structures (green) aligned to holo structures (gray).
(C) Fast-folding proteins from DESRES. DyneTrion recovers MD-like free-energy landscapes and metastable-state populations under leave-one-out fine-tuning, as assessed by time-lagged independent component analysis (TICA) projections and ensemble distributions.
  }
  \label{fig:MD_emulation_extrapolation}
\end{figure}

To this end, we evaluate DyneTrion along three complementary axes (Fig.~\ref{fig:MD_emulation_extrapolation}). First, we assess whether DyneTrion preserves the free-energy landscape and dynamical observables of microsecond MD simulations. Second, we test whether the model can propagate proteins between experimentally observed conformational states, focusing on apo-to-holo transitions. Finally, we consider an extreme regime using fast-folding proteins, where folding events occur on hundreds-of-microseconds to millisecond timescales, requiring both long-range extrapolation and correct ensemble shaping.

\paragraph{Microsecond-Scale Ensemble Fidelity}
We first evaluated DyneTrion on microsecond-scale simulations from dynamicPDB (Fig.~\ref{fig:MD_emulation_extrapolation}A), comprising 20 proteins with trajectories extending up to 1~$\mu$s. These systems provide a stringent test of long-timescale extrapolation, as accurate generation requires recovering both the correct conformational ensemble and the associated free-energy landscape between partially unfolded and folded states. DyneTrion achieves a mean absolute error (MAE) of 0.51~kcal/mol and a Pearson correlation of 0.79 with MD-derived free energies across test proteins.

Beyond free-energy accuracy, Fig.~\ref{fig:MD_emulation_extrapolation}A (right) shows that key dynamical observables remain stable as the extrapolation length increases from 100~ns to 1~$\mu$s. Measures of flexibility, including pairwise C$\alpha$–RMSD and global and per-target RMSF, show minimal drift across timescales, while ensemble observables such as weak contacts, transient contacts, and solvent-exposed residues remain consistent. DyneTrion also maintains stable structural quality across extrapolation horizons, with no systematic degradation in stereochemical validity or structure coverage as the trajectory length increases. Together, these results demonstrate that DyneTrion preserves both thermodynamic and dynamical fidelity during microsecond-scale extrapolation, rather than exhibiting progressive degradation with increasing trajectory length.

\paragraph{Apo-to-Holo Conformational Propagation}
To evaluate whether DyneTrion can propagate proteins between distinct functional states, we examined apo-to-holo transitions for 72 proteins curated in by Jing, \textit{et.al.}~\cite{jing2023eigenfold} (Fig.~\ref{fig:MD_emulation_extrapolation}B). Starting from experimentally determined apo conformations, DyneTrion generates structures that move toward the corresponding holo states, achieving an average maximin TM-score of approximately 0.83. 

As shown in Fig.~\ref{fig:MD_emulation_extrapolation}B (left), holo-like conformations are indeed found in DyneTrion’s generated trajectories, yielding structures with maximin TM-scores that are closer to the holo state than to the starting apo state—even for proteins with substantial apo–holo differences (color-coded by intrinsic apo–holo TM-score). Representative examples (Fig.~\ref{fig:MD_emulation_extrapolation}B, right) illustrate that DyneTrion captures coordinated domain and secondary-structure rearrangements, yielding RMSDs of 1.29~\AA{} and 2.00~\AA{} relative to experimental holo structures for antibody and membrane protein systems, respectively. Together, these results indicate that DyneTrion can traverse large-scale conformational changes and generate physically plausible intermediate-to-holo-like structures, rather than merely sampling local fluctuations around the apo input.

\paragraph{Fast-Folding Proteins under Leave-One-Out Generalization}
Finally, we evaluated DyneTrion in an extreme extrapolation regime using 12 fast-folding proteins simulated on the Anton supercomputer by D.~E.~Shaw Research~\cite{lindorff2011fast}. These systems are intentionally challenging, as folding events typically occur on hundreds-of-microseconds to millisecond timescales. Following the evaluation protocol of Lewis, \textit{et al.}~\cite{bioemu2025}, we performed leave-one-out cross-validation, fine-tuning DyneTrion on 11 proteins and evaluating on the held-out target (Fig.~\ref{fig:MD_emulation_extrapolation}C).

As shown by the TICA projections in Fig.~\ref{fig:MD_emulation_extrapolation}C, DyneTrion recovers free-energy surfaces that closely match those obtained from MD simulations, including folded, unfolded, and intermediate basins. Quantitatively, the Jensen–Shannon divergence between model-generated and MD TICA distributions is approximately 0.5 across test proteins, indicating substantial overlap of the sampled conformational ensembles. For representative systems such as Villin and Protein G, both MD and DyneTrion sample comparable intermediate states, including partially unfolded helices while preserving $\beta$-sheet structure, consistent with known folding pathways. These results demonstrate that DyneTrion generalizes beyond the training set to capture rare folding-related conformational transitions, despite the long intrinsic timescales required to observe such events in MD.

\subsection{Ensembles Generation}

\begin{figure}[t!]
  \centering
  \includegraphics[width=1.0\linewidth]{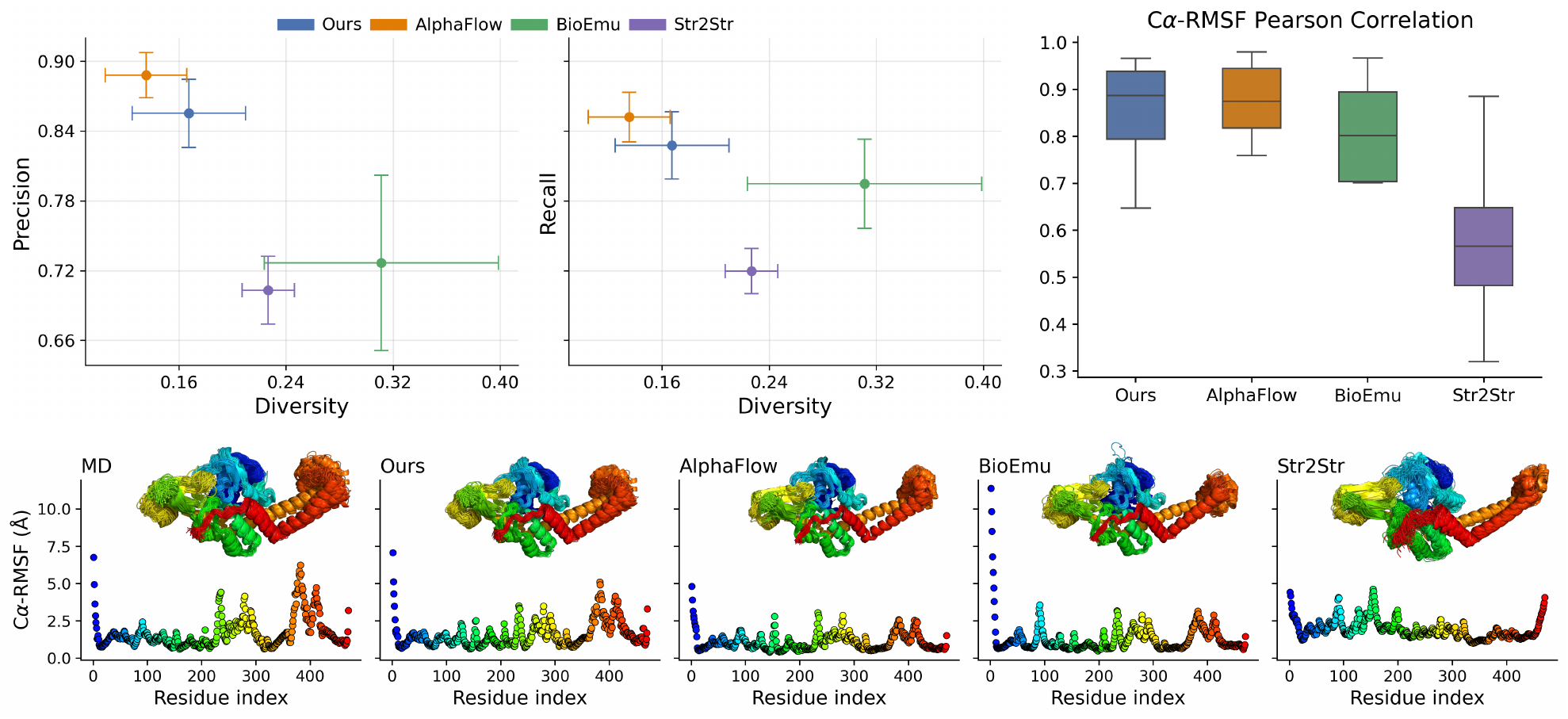} %
  \caption{\textbf{Comparison with other generative model for protein ensembles.} The precision, recall and diversity for benchmarked methods, and the pearson correlation between generated structures and MD simulations based on C$\alpha$-RMSF. Insets show ensembles of PDB ID 3ilw(Chain A) with the C$\alpha$-RMSF plotted by residue index. 
  }
  \label{fig:conformation_sampling}
\end{figure}

We further compared DyneTrion with established generative ensemble-sampling methods to assess its ability to generate diverse and physically meaningful protein conformational ensembles over long timescales. We evaluated 20 proteins from dynamicPDB, each associated with 1~$\mu$s MD simulations. For each protein, we  sampled 64 structures with the highest structural difference from the generated trajectories, and all methods were evaluated using the same number of samples to ensure a fair comparison. Baseline methods include AlphaFlow~\cite{jing2023alphaflow}, BioEmu~\cite{bioemu2025}, and Str2Str~\cite{lu2024strstr}, with results summarized in Fig.~\ref{fig:conformation_sampling}.

As shown in Fig.~\ref{fig:conformation_sampling} (top), DyneTrion achieves a favorable balance between ensemble diversity and structural accuracy, attaining precision and recall comparable to AlphaFlow, which is explicitly designed for ensemble generation. Notably, DyneTrion exhibits competitive diversity while maintaining high precision, indicating that its generated ensembles cover a broad conformational space without sacrificing structural relevance. In contrast, Str2Str and BioEmu exhibit higher diversity at the cost of reduced precision and recall, reflecting different trade-offs in ensemble construction.

Agreement with MD-derived flexibility is further assessed via residue-level C$\alpha$-RMSF correlations (Fig.~\ref{fig:conformation_sampling}, top right). DyneTrion achieves a median Pearson correlation of approximately 0.90, comparable to AlphaFlow ($\sim$0.87) and higher than BioEmu and Str2Str, indicating accurate recovery of residue-resolved fluctuation patterns from ensemble statistics. Representative RMSF profiles (Fig.~\ref{fig:conformation_sampling}, bottom) show that DyneTrion closely reproduces both the magnitude and spatial distribution of flexibility observed in MD, including highly mobile loop regions and more rigid secondary-structure elements.

Together, these results demonstrate that DyneTrion functions not only as a trajectory generator but also as an effective ensemble generator, achieving performance on par with specialized ensemble-sampling models while preserving the dynamical signatures encoded in long-timescale MD simulations.

\section{Discussion}\label{sec3}

We introduce DyneTrion, a 4D diffusion framework for emulating protein dynamics directly from molecular dynamics (MD) data. By representing protein motion as time-evolving rigid-body transformations and integrating explicit spatial–temporal attention, DyneTrion generates physically plausible protein dynamics across multiple timescales, moving beyond static structure prediction toward dynamic ensemble modeling.

On short timescales, DyneTrion accurately reproduces residue-level flexibility patterns observed in MD simulations, achieving strong correlations in pairwise RMSD and RMSF. Importantly, the model maintains stereochemical validity during extrapolation, avoiding the rapid accumulation of clashes and backbone breaks seen in offset-based trajectory models. These results indicate that DyneTrion learns structured dynamical correlations rather than fitting framewise coordinate noise.

DyneTrion further demonstrates coherent extrapolation to microsecond regimes, preserving ensemble-level free energies and dynamical observables as the simulation horizon increases. Long-timescale evaluations show stable recovery of flexibility statistics, contact patterns, and solvent exposure. In challenging fast-folding systems, the model recovers MD-like free-energy landscapes and known metastable intermediates under leave-one-out generalization, while apo-to-holo experiments show that DyneTrion can propagate proteins between distinct functional conformations.

Despite these strengths, DyneTrion currently learns dynamics implicitly without force or energy supervision and is trained on single-chain proteins under fixed thermodynamic conditions. Addressing these limitations, together with developing principled uncertainty estimates for ensemble observables, will be important directions for future work. Overall, DyneTrion highlights the potential of diffusion-based generative models with geometric and temporal inductive biases to bridge the gap between static structure prediction and full molecular dynamics simulations.

\section{Methods}\label{sec4}
\subsection{Preliminaries}\label{sec:preliminary}
\paragraph{Protein Parameterization.}
We employed the frame-based representation of protein structure used in AlphaFold2~\cite{jumper2021highly} and extended it to incorporate a temporal dimension, accounting for structural changes over time.
A static protein is represented as a sequence of amino acid residues, each parameterized by a backbone frame and side-chain. The backbone consisting of atoms [$\mathtt{N}, \mathtt{C}_{\alpha}, \mathtt{C}$] with $\mathtt{C}_{\alpha}$ positioned at the residue origin $(0, 0, 0)$. 
We defined the protein dynamic trajectory composed of $N$ amino acid residues, each parameterized by a backbone frame that transforms $M$ time steps. 
Those frames can be transformed by special Euclidean transformations, mapping local frames to a global coordination system, represented by $T_{m,n}=[R_{m,n}, X_{m,n}]\in\mathrm{SE}(3)$, where $m\in\{1,...,M\}$, $n\in\{1,...,N\}$, $R_{m,n}\in\mathrm{SO}(3)$ is a $3\times3$ rotation matrix, and $X_{m,s}\in\mathbb{R}^3$ is the translation vector.
The coordinates of sidechains in the residue are grouped into rigids units based on the torsion angles with respected to the backbone. This setup allows each residue to be parameterized by torsion angles $\alpha_{m,n}\in\mathbb{R}^7$, which model the rotations required to align atom groups relative to the backbone.
The torsion angles enable precise adjustments of atom positions within each frame, while the transformation parameters allow the model to reconstruct all atom positions from idealized, experimentally determined coordinates over time.

\paragraph{Score-based Modeling on $\mathrm{SE}(3)^{M\times N}$.}
The score-based generative model estimates the gradient of the data distribution, referred to as the score function~\cite{song2019generative}, by progressively perturbing the data into a noise distribution via a stochastic differential equation (SDE), and then generating samples by solving the corresponding reverse-time SDE~\cite{anderson1982reverse} with learned score function.

Following prior work~\cite{yim2023se}, we represent a protein trajectory using frame representations~\cite{jumper2021highly} $T = [T_{m,n}] \in \mathrm{SE}(3)^{M \times N}$, where $M$ denotes the number of structures and $N$ the number of residues, and apply a diffusion process that gradually transforms these frames from noise to plausible structures.
More specifically, we construct two independent forward processes for $R=[R_{m,n}]\in\mathrm{SO}(3)^{M\times N}$ and $X=[X_{m,n}]\in\mathbb{R}^{M\times N\times 3}$ respectively: 
\begin{align}
\mathrm{d}T^{(t)} &= [\mathrm{d}R^{(t)}, \mathrm{d}X^{(t)}] \notag\\
                  &= \left[0, -\frac{1}{2}X^{(t)}\right]\mathrm{d}t + [\mathrm{d}B^{(t)}_{\mathrm{SO}(3)^{M\times N}}, \mathrm{d}B^{(t)}_{\mathbb{R}^{M \times N \times 3}}],
\end{align}
where $B^{(t)}_{\mathrm{SO}(3)^{M\times N}}$ and $B^{(t)}_{\mathbb{R}^{M\times S\times 3}}$ are the Brownian motion on $\mathrm{SO}(3)^{M\times N}$ and $\mathbb{R}^{M\times N\times 3}$, and $t\in[0, 1]$ denotes the diffusion time steps. 
Superscripts in parentheses are used to represent specific time steps. 
Lowercase letters denote deterministic variables, and uppercase letters denote random variables.

Accordingly, the associated backward process is given by the equation $\mathrm{d}\overleftarrow{T}^{(t)}=[\mathrm{d}\overleftarrow{R}^{(t)}, \mathrm{d}\overleftarrow{X}^{(t)}]$, where
\begin{align}
    \mathrm{d}\overleftarrow{R}^{(t)} &= \nabla\log p_{1-t}(\overleftarrow{R}^{(t)})\mathrm{d}t + \mathrm{d}B^{(t)}_{\mathrm{SO}(3)^{M\times N}}, \\ 
    \mathrm{d}\overleftarrow{X}^{(t)} &= (\frac{1}{2}\overleftarrow{X}^{(t)}+\nabla\log p_{1-t}(\overleftarrow{X}^{(t)}))\mathrm{d}t + \mathrm{d}B^{(t)}_{\mathbb{R}^{M\times N\times 3}}.
\end{align}
The score function at diffusion time $t$ is then given by
\begin{equation}
\nabla \log p_t(T^{(t)})=[\nabla \log p_t(R^{(t)}), \nabla \log p_t(X^{(t)})]
\end{equation}
and is approximated using a neural networks $s_{\theta}(t, T^{(t)})$ trained by minimizing the denoising score matching(DSM) loss:
\begin{equation}
    \mathcal{L}(\theta) = \mathbb{E}[\lambda_t||\nabla \log p_{t|0}(T^{(t)}|T^{(0)})-s_{\theta}(t, T^{(t)})||^2],
    \label{eq:dsm}
\end{equation}
where  the expectation is taken over $t\sim\mathcal{U}[0,1]$, and the conditional score functions at time step $t$ is:
\begin{equation}
\nabla \log p_{t|0}(T^{(t)}|T^{(0)}) = \begin{aligned}[t]
&[\nabla \log p_{t|0}(R^{(t)}|R^{(0)}), \\
&\nabla \log p_{t|0}(X^{(t)}|X^{(0)})].
\end{aligned}
\end{equation}
The $\lambda_t\in\mathbb{R}^+$ is a diffusion-time dependent weighting factor.

\subsection{Model Architecture.}
The architecture of our model is depicted in Fig.~\ref{fig:pipeline}, which contains a protein sequence encoding and a diffusion model.  A protein dynamic trajectory is represented by $N$ residues across $M$ structures. We first use multiple sequence alignment(MSA), Embeder and Evoformer from AlphaFlod2~\cite{jumper2021highly} to build single $\textbf{s}$ and pair representation $\textbf{z}$ for the input amino acid sequence. We use MLP to embed the diffusion time step $t$ and add to single and pair presentations. Based on these representations, we employ a invariant point attention layer to incorporate the noise rigids, reference structure and motion information into the representations. Specifically, we represent the feature at at layer $l\in \{1,...,L\}$ using node representations $\mathbf{S}^l=[\mathbf{s}^l_{m,n}]\in \mathbb{R}^{M\times N\times c_s}$ and edge representations $\mathbf{Z}^l=[\mathbf{z}^l_{m,(n,k)}]\in \mathbb{R}^{M\times N\times N\times c_z}$. Here, $\mathbf{S}^l$ denotes the feature of the residue $n$ in layer $l$ and $\mathbf{Z}^l$ encodes the relationship between residue $n$ and $k$ for the $m$-th structure in the trajectory. The positional attribution of residues are represented by rigid transformation $\mathbf{T}=[T_{m,n}]\in \mathbb{R}^{M\times N \times 6}$. 
The reference structure is defined by $\mathbf{S}^{l}_{\mathtt{ref}}$, $\mathbf{Z}^{l}_{\mathtt{ref}}$, and $\mathbf{T}^{l}_{\mathtt{ref}}$. 
Motion structures are characterized by $\mathbf{S}^{l}_{\mathtt{mot}}$, $\mathbf{Z}^{l}_{\mathtt{mot}}$, and $\mathbf{T}^{l}_{\mathtt{mot}}$, which  describe multi-order motion information, including velocity, acceleration, and other related parameters. Besides, we use a MLP to generated torsion angles $\mathbf{\alpha}=[\alpha_{m,n}]$ based on the node representations from layer $L$.

\paragraph{Feature Embedding from Amino Acid Sequence.}
We initially extract single and pair features for a given amino acid sequence following AlphaFold2~\cite{jumper2021highly}. 
To incorporate the diffusion time step $t$, we project $t$ into time embeddings 
$[t_{\mathrm{node}},\, t_{\mathrm{edge}}]$ 
using independent MLPs, which are then added to the single and pair representations as
$\mathbf{s} \leftarrow \mathbf{s} + t_{\mathrm{node}}$ and 
$\mathbf{z} \leftarrow \mathbf{z} + t_{\mathrm{edge}}$, respectively.

\paragraph{Iterative Update.}
The iterative update process is performed across each network layer $l$, where node representations are processed first, followed by edge representations and rigids transformation. Specifically, at layer $l$, the node representations $\mathbf{S}^{l+1}_{m,n}$ and $\mathbf{S}^{l+1}_{m,k}$ for residues $n$ and $k$ are concatenated and and added to the edge feature $\mathbf{Z}^{l}_{m,(n,k)}$. The resulting features are processed with a MLP layer to produce edge representations $\mathbf{Z}^{l+1}_{m,(n,k)}$.
Simultaneously, a structural update $\Delta T^l_{m,n}$ is computed from the nodes $\mathbf{S}^{l}_{m,n}$ for each residue $n$ via linear projection layers and applied to the current frame to obtain the updated rigids transformation $T^{l+1}_{m,n}$. This iterative procedure of updating node representations, edge representations, and rigids is repeated throughout network layers.

\paragraph{Spatial-Temporal Transformer}
The Spatial-Temporal Transformer is designed to jointly model spatial structural relationships and temporal coherence in dynamic protein trajectories. Initially, we integrate the node representations $\mathbf{S}^l$, edge representations $\mathbf{Z}^l$ and the corresponding rigids $\mathbf{T}$ for the reference, motion and noisy structures through the IPA (Fig.~\ref{fig:pipeline} B). We then replicate reference node representation $\mathbf{S}^l_{\mathrm{ref}}\in \mathbb{R}^{1\times N \times c_s}$ along the temporal dimension to obtain $\mathbf{\tilde{S}}^l_{\mathrm{ref}} \in \mathbb{R}^{M\times N \times c_s}$. This expanded presentation is concatenated with $\mathbf{S}^l$ along the feature dimension, forming $[\mathbf{S}^l,\mathbf{\tilde{S}}^l]\in \mathbb{R}^{M\times N \times 2c_s}$, which is subsequently processed by a transformer block~\cite{vaswani2017attention} to process the combined representations and obtain the output node representation $\mathbf{\tilde{S}}^l$.
To model the protein's dynamic behavior, we first transpose the motion representations $\mathbf{S}^l_{\mathrm{mot}}\in \mathbb{R}^{M_{\mathtt{mot}}\times N \times c_s}$ into $\mathbb{R}^{ N \times M_{\mathtt{mot}}\times c_s}$. Likewise,  the representations $\mathbf{\tilde{S}}^l\in \mathbb{R}^{ N \times M\times c_s}$ are aligned accordingly. These resulting representation are then concatenated along the temporal dimension to form $[\mathbf{S}^l_{\mathrm{mot}},\mathbf{\tilde{S}}^l] \in \mathbb{R}^{ N \times (M_{\mathrm{mot}}+M)\times c_s}$. Temporal sinusoidal positional embedding~\cite{vaswani2017attention} are added to this combined representations, which are subsequently processed by a transformer layer.

\subsection{Training Loss}

The score model is trained end-to-end using gradient from the primary DSM loss, torsion angles loss and several auxiliary losses. The total per-example loss is defined as:
\begin{equation}
\mathcal{L}=\mathcal{L}_{\mathtt{dsm}}+w_1\cdot \mathds{1}\{t<\frac{1}{4}\}(\mathcal{L}_{\Omega}+\mathcal{L}_{2D})+w_2\cdot \mathcal{L}_{torsion},
\end{equation}
where $w_1$ and $w_2$ denote the weighting coefficients, we set $w_1=0.25$ and $w_2=1$ in our experiments. $\mathds{1}(\cdot)$ is the indicator function, which evaluates to 1 when the condition is satisified and 0 otherwise.

\paragraph{Denoising Score Matching Loss.} 
The neural network is trained to learn the rotation and translation score function by minimizing Equation~\ref{eq:dsm}.
Specifically, we adopt a diffusion-time dependent weight schedule for the rotation component,
\begin{equation}
\lambda_t^R=1/\mathbb{E}[||\nabla \log p_{t|0}(R^{(t)}|R^{(0)})||^2_{\mathrm{SO}(3)}].
\end{equation}
For the translation component, we use
\begin{equation}
\lambda_t^X=(1-\exp^{-t})/\exp^{-\frac{t}{2}}
\end{equation} to prevent instability in loss values at low $t$.
The DSM loss is defined as follows:
\begin{equation}
\mathcal{L}_{\mathtt{dsm}}=\mathcal{L}_{\mathtt{dsm}}^{R} + \mathcal{L}_{\mathtt{dsm}}^{X}.
\end{equation}

\paragraph{Torsion Angle Loss.}
We employ a multi-layer perceptron (MLP) to predict the side chain and backbone torsion angles~\cite{jumper2021highly}, $\mathbf{\alpha}=[\alpha_{m,n}]$, based on the node representations from layer $L$. The predicted torsion angles are normalized and represented on the unit circle, with $\Vert \alpha_{m,n} \Vert \in\mathbb{R}^{7\times 2}$ for sine and cosine values.
Due to the $180^{\circ}$ rotational symmetry of some side chains, the model is permitted to predict either the ground truth torsion angles or an alternative set of angles:
\begin{equation}
\mathcal{L}_{\mathtt{torsion}}= \frac{1}{M}\frac{1}{N}\sum_{m=1}^{M}\sum_{n=1}^{N} (\min (\Vert \alpha_{m,n}-\alpha_{m,n}^{\mathrm{gt}} \Vert ^2, \Vert \alpha_{m,n}-\alpha_{m,n}^{\mathrm{alt,gt}} \Vert ^2 ))
\end{equation}
where $\alpha_{m,n}$,$\alpha_{m,n}^{\mathrm{gt}}$ and $\alpha_{m,n}^{\mathrm{alt,gt}}$ represent predicted, ground truth, and alternative ground truth torsion angles, respectively, for residue $n$ in $m$-th structure in trajectory.

\paragraph{Auxiliary loss.} 
To mitigate chain breaks and steric clashes, we impose axiliary penalties on atomic-level error. We define the backbone atoms as $\Omega=\{\mathtt{N}, \mathtt{C}, \mathtt{C}_{\alpha}, \mathtt{O}\}$ and compute the mean squared error on the positions of selected atoms in $\Omega$: 
\begin{equation}
\mathcal{L}_{\mathtt{\Omega}}=\frac{1}{M}\frac{1}{4N}\sum_{m=1}^{M}\sum_{n=1}^{N}\sum_{a_i\in\Omega_{m,n}} ||a_i-\hat{a}_i||^2
\end{equation}
where $a_i$ and $\hat{a}_i$ are the ground-truth and predicted atom positions from $\Omega_{m,n}$, respectively.
In addition, we introduce a second auxiliary loss penalizes pairwise atomic distance errors:
\begin{equation}
\mathcal{L}_{\mathtt{2D}}=\frac{1}{C}\sum_{m=1}^{M}\sum_{n=1}^N\sum_{a_i,b_j\in\Omega_{m,n}}\mathds{1}\{d_{ab}^{ij}<0.6\}||d_{ab}^{ij}-\hat{d}_{ab}^{ij}||^2
\end{equation}
where $d_{ab}^{ij}=||a_{i}-b_{j}||$ denotes the pairwise atomic distance, and $C=\sum_{m=1}^M(\sum_{n=1}^N\sum_{a_i,b_j\in\Omega_{m,n}}\mathds{1}\{d_{ab}^{ij}<0.6\}-N)$. The indicator function $\mathds{1}\{d_{ab}^{ij}<0.6\}$  ensures only atom pairs within 0.6 nanometer are penalized.

\subsection{Molecular Simulation}

\subsubsection{Source of Protein Structural Data.}
Currently, there are experimentally determined protein structures deposited in the Protein Data Bank (PDB)~\cite{berman2002protein}. 
Despite the extensive scope of the PDB database, some structures, such as membrane proteins, present challenges for molecular dynamics (MD) simulations~\cite{lindahl2008membrane}. 
To address these challenges, we applied a series of preprocessing steps to prepare the structures for MD simulations. We adhere to a systematic approach involving selection, cleaning, and completion to prepare proteins for molecular dynamics simulations. 

\paragraph{Selection.} 
Initially, we select proteins with structures determined via X-ray diffraction, ensuring a resolution of no greater than 2.0 \AA. 
Subsequently, we filter out monomeric single-chain proteins in their oligomeric state and possess a sequence length of 500 residues or fewer. 
We further exclude membrane proteins by consulting databases such as OPM~\cite{lomize2006opm}, PDBTM~\cite{kozma2012pdbtm}, MemProtMD~\cite{newport2019memprotmd}, and mpstruc~\cite{bittrich2022rcsb}. 
Lastly, we utilize the dictionary of secondary structure in proteins (DSSP)~\cite{kabsch1983dictionary} to eliminate proteins characterized by more than 50\% coil-loop conformations.

\paragraph{Cleaning.} 
In this step, we remove all heteroatoms from the protein structures, including water molecules and ligands, to focus on the intrinsic dynamic behavior of the proteins. Additionally, we standardize the nomenclature by replacing non-standard residue names with their corresponding standard designations.

\paragraph{Completion.} 
For proteins exhibiting five or fewer missing residues, we employ MODELLER~\cite{webb2016comparative} for reconstruction. 
Conversely, for proteins with six or more missing residues, restoration is achieved utilizing AlphaFold2~\cite{jumper2021highly}. 
Furthermore, we incorporate hydrogen atoms into the protein structures using MODELLER to ensure completeness.

\subsection{Implement Detail}

\paragraph{Molecular Simulation}

All-atom molecular dynamics simulations are conducted using OpenMM~\cite{eastman17openmm} version 8.0.0 in conjunction with the Amber-ff14SB force field, which has been shown to enhance the accuracy of protein side chain and backbone parameters.
The dimensions of the periodic box containing each protein are defined by a padding distance of 1 nm. 
This box is filled with TIP3P water molecules and subsequently neutralized with $\text{Na}^+/\text{Cl}^-$ ions at a concentration of 150 $mM$.

To mitigate structural artifacts, minimize atomic clashes, and establish a stable starting conformation for reliable simulation outcomes, an energy minimization process is performed. 
The force tolerance for this minimization is set to $2.39 \,kcal/mol \cdot nm$ without any imposed maximal step limits. 
Following energy minimization, two sequential equilibration processes are carried out: first in a canonical ensemble (NVT) and subsequently in an isothermal-isobaric ensemble (NPT), each spanning 1 ns with a time step of 1 fs. 
The \textit{Langevin~Middle~Integrator} is employed as the integrator for both equilibration phases, with the heat bath temperature and friction coefficient set to 300 $K$ and 1.0 ps$^{-1}$, respectively. 
During the NPT equilibration phase, the pressure is maintained at 1.0 bar utilizing the Monte Carlo Barostat.

Post-equilibration, the primary molecular dynamics simulations are executed, with each protein being simulated for a duration of 1 $\mu$s. 
A short step of 1 fs ensures computational stability throughout the simulation process. 
Atomic coordinates and various physical properties, including energy, velocity, and force, are recorded every 1 ps, yielding a total of 1.0M data frames.

The simulations were performed on the GPU platform utilizing multiple processes, each running on a 16-core Intel Xeon CPU operating at 2.90 GHz and supported by an NVIDIA A100 GPU with 80 GB of memory. 
The raw data generated from the MD trajectories and associated physical properties for the selected proteins form the foundation of our dataset.

\begin{figure}[t!]
  \centering
  \includegraphics[width=1.0\linewidth]{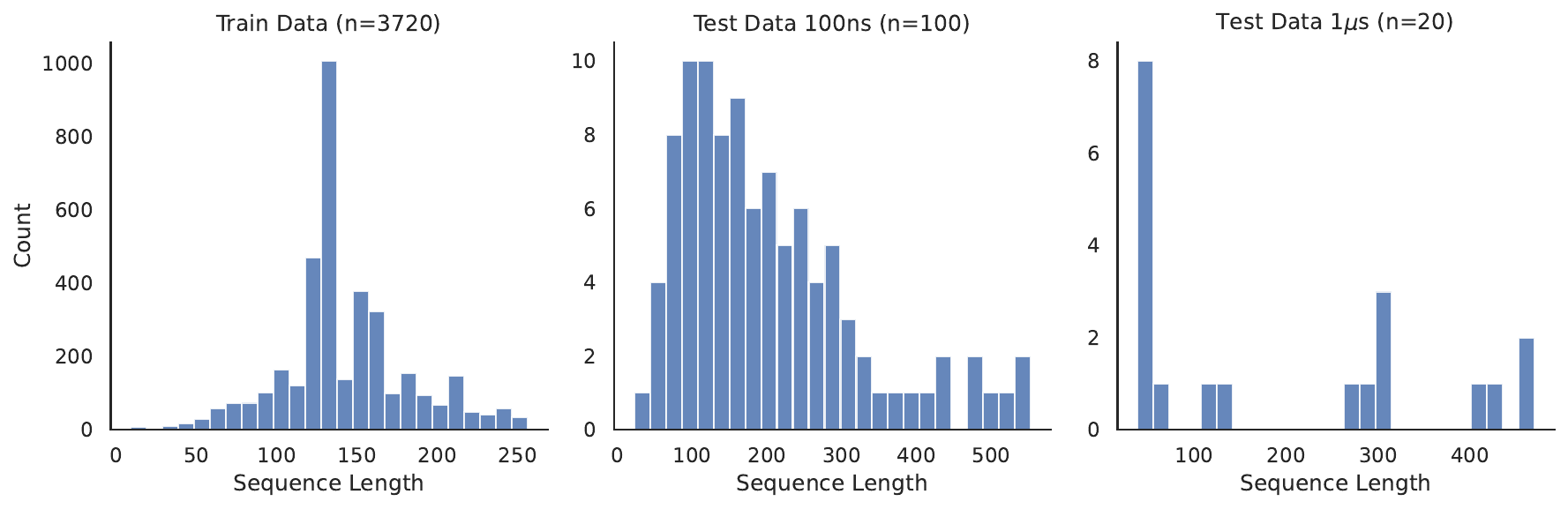}
  \caption{Histogram of sequence lengths in the train and test data sets}
  \label{fig:data_distribution}
\end{figure}

\paragraph{Data Curation}
Our dataset comprises approximately 4.0k protein dynamic trajectories curated from a diverse range of organisms using the May 2024 release of the Protein Data Bank (PDB), with each protein containing fewer than 500 residues. Proteins were first clustered at 40\% sequence identity using MMseqs2~\cite{steinegger2017mmseqs2}, and the train/test split was performed at the cluster level to prevent information leakage. The final test set consists of 100 proteins randomly sampled from the held-out clusters. Due to GPU memory constraints, only proteins with fewer than 256 residues were used for training, resulting in approximately 3.7k training trajectories. The dataset distribution is illustrated in Fig.~\ref{fig:data_distribution}.

\paragraph{Training Hyperparameters}
We trained the diffusion model using 100 denoising steps. The model comprises four blocks, each containing an IPA module and an EdgeUpdate module, with a hidden dimension of 256. The proposed score model includes approximately 23.07 million trainable parameters (Fig.~\ref{fig:pipeline}B). Training was conducted using the Adam optimizer with a learning rate of $1\times10^{-4}$ and a batch size of 4. The model was trained for 450 epochs over approximately 48 hours on four NVIDIA A100 GPUs, generating 16 structures with 400ps intervals per inference. We adopt an SE(3)-equivariant diffusion framework and perform residue-level SDE-based denoising separately for translations and rotations. Specifically, a variance-preserving SDE (VP-SDE) is applied to translations with $\beta_{\min}=0.1$ and $\beta_{\max}=20$, while a variance-exploding SDE (VE-SDE) is applied to rotations with $\sigma_{\min}=0.1$ and $\sigma_{\max}=1.5$, following~\cite{yim23se3}.

\paragraph{Baseline Model Implementation}
We compare our method with several generative models for protein dynamic trajectories, including MDGEN~\cite{jing2024generative}, UniSim~\cite{yu2025unisim}, and ProTDyn~\cite{liu2025protdyn}. For a fair comparison, MDGEN and UniSim are retrained on the same training dataset as our model. For ProTDyn, we use the publicly available pretrained open-source version with a simulation resolution of 1$\sim$ns. All models are implemented to extrapolate to the same timescales as the MD simulations.

For protein ensemble generation, we evaluated baseline methods using their publicly released pretrained models. Specifically, we employ AlphaFlow~\cite{jing2023alphaflow} with template conditioning and 10 denoising steps, BioEmu~\cite{bioemu2025} v1.1 without the unphysical filter, and Str2Str~\cite{lu2024strstr} with  $\Delta t = 0.2$. All models are evaluated by generating 64 ensemble structures for comparison.

\subsection{Evaluation Metrics}\label{sec:metric}
\subsubsection{Dynamic Property} We adopt statistic metric from AlphaFlow~\cite{jing2023alphaflow} to evaluate the short time-scale results under three categories: (1) predicting flexibility, (2) distribution accuracy, and (3) ensemble observables.

\paragraph{Predicting flexibility} Predicting flexibility is assessed by the C$\alpha$-RMSD between any pair of structures with each trajectory and quantifying the Pearson correlation with the MD results, reported as the the Pairwise RMSD $r$. Atomic-level flexibility is characterized by the root mean square fluctuation(RMSF), computed in an analogous manner. RMSF values are then aggregated both globally and on a per-target basis to yield the Global RMSF and Per-target RMSF metrics, respectively.

\textbf{Pairwise RMSD $r$:} let a trajectory consist of $M$ conformations. For conformation $m$,  $\mathbf{X}_m \in \mathbb{R}^{N \times 3}$ denote the C$\alpha$ coordinates of $N$ atoms.  
The pairwise C$\alpha$-RMSD between two conformations $m$ and $m'$ in a trajectory
is defined as
\begin{equation}
\mathrm{RMSD}^{\mathrm{C}\alpha}_{m,m'}
=
\sqrt{
\frac{1}{N}
\sum_{n=1}^{N}
\left\|
\mathbf{x}^{\mathrm{C}\alpha}_{m,n}
-
\mathbf{x}^{\mathrm{C}\alpha}_{m',n}
\right\|_2^2
},
\qquad 1 \le m < m' \le M ,
\end{equation}
where $M$ is the number of conformations in the trajectory and $N$ is the number
of C$\alpha$ atoms. We then summarize pairwise distances within a trajectory by averaging over all
distinct conformation pairs:
\begin{equation}
\mathrm{Pairwise\ RMSD}
=
\frac{2}{M(M-1)}
\sum_{1 \le m < m' \le M}
\mathrm{RMSD}^{\mathrm{C}\alpha}_{m,m'} .
\end{equation}
All trajectory-level Pairwise RMSD values are collected across test proteins and
compared with those computed from MD trajectories using the Pearson correlation
coefficient, reported as the Pairwise RMSD $r$:
\begin{equation}
r_{\mathrm{Pairwise\ RMSD}}
=
\mathrm{corr}
\left(
\{ \mathrm{Pairwise\ RMSD}^{\mathrm{gen}}_{j} \},
\{ \mathrm{Pairwise\ RMSD}^{\mathrm{MD}}_{j} \}
\right),
\end{equation}
where $j$ indexes test proteins.

\textbf{Global RMSF $r$ and Per-target RMSF $r$:}  we define the C$\alpha$ RMSF of residue $n$ as
\begin{equation}
\mathrm{RMSF}_n
=
\sqrt{
\frac{1}{M}
\sum_{m=1}^{M}
\left\|
\mathbf{x}_{m,n}
-
{\mathbf{x}}_n^\mathtt{ref}
\right\|^2
}.
\end{equation}
where $\mathbf{x}^{\mathrm{ref}}_{n}$ denotes the C$\alpha$ coordinate of residue $n$ in the reference structure obtained from the Protein Data Bank (PDB).
The Global RMSF $r$ is computed as the Pearson correlation between generated and MD RMSF values, aggregated over all residues across all test proteins. Let $j$ index protein targets and $n$ index residues within each protein:
\begin{equation}
r_{\mathrm{Global\ RMSF}}
=
\mathrm{corr}
\left(
\left\{ \mathrm{RMSF}^{\mathrm{gen}}_{j,n} \right\}_{j,n},
\left\{ \mathrm{RMSF}^{\mathrm{MD}}_{j,n} \right\}_{j,n}
\right).
\end{equation}
For each protein target $j$, we compute an RMSF correlation:
\begin{equation}
r_{\mathrm{RMSF}}^{(j)}
=
\mathrm{corr}
\left(
\{ \mathrm{RMSF}^{\mathrm{gen}}_{j,n} \}_{n},
\{ \mathrm{RMSF}^{\mathrm{MD}}_{j,n} \}_{n}
\right).
\end{equation}
The Per-target RMSF $r$ is reported as the median over all $J$ test
proteins:

\paragraph{Distribution accuracy} Distribution accuracy is evaluated using the root mean Wasserstein distance(RMWD) between generated trajectories and MD trajectories. In addition, we project the C$\alpha$ positions onto the first two principal components from principal component analysis(PCA). We first fit the PCA using the MD trajectories, transform the generated trajectories accordingly, and compute the Wasserstein distance between MD and generated projections, reported as MD PCA $\mathcal{W}_2$. We further perform a joint PCA fit on both MD and generated trajectories, and compute the   Wasserstein distance denoted as the Joint PCA $\mathcal{W}_2$. PCA is performed separately on the MD and the generated, and the cosine similarity  between components is computed for PC-sim. 

\textbf{Root Mean Wasserstein Distance (RMWD).} Given a generated trajectory ensemble $\mathcal{X}$ and a corresponding molecular dynamics (MD) ensemble $\mathcal{Y}$, we define the root mean Wasserstein distance (RMWD) as:
\begin{equation}
\begin{aligned}
\mathrm{RMWD}^2(\mathcal{X}, \mathcal{Y})
= {} &
\frac{1}{N} \sum_{i=1}^{N}
\left\|
\boldsymbol{\mu}_{\mathcal{X},i}
-
\boldsymbol{\mu}_{\mathcal{Y},i}
\right\|_2^2
\\
& +
\frac{1}{N} \sum_{i=1}^{N}
\mathrm{Tr}
\Bigl(
\boldsymbol{\Sigma}_{\mathcal{X},i}
+
\boldsymbol{\Sigma}_{\mathcal{Y},i}
-
2
\bigl(
\boldsymbol{\Sigma}_{\mathcal{X},i}^{1/2}
\boldsymbol{\Sigma}_{\mathcal{Y},i}
\boldsymbol{\Sigma}_{\mathcal{X},i}^{1/2}
\bigr)^{1/2}
\Bigr).
\end{aligned}
\end{equation}
Here, $i$ indexes atoms, while $\boldsymbol{\mu}_{\mathcal{X},i}$ and $\boldsymbol{\Sigma}_{\mathcal{X},i}$ denote the empirical mean position and covariance matrix of atom $i$ estimated from $\mathcal{X}$, respectively (and analogously for $\mathcal{Y}$). The first term quantifies the translational discrepancy between the two trajectories, whereas the second term captures differences in atomic positional variability through their covariance structure. The RMWD values are summarized by their median across all test proteins.

\textbf{PCA-based Distribution Comparison.}
 We first fit principal component analysis (PCA) on the MD trajectories, denoted as $\mathcal{P}_{\mathtt{md}}$. Analogously, PCA is performed on the generated trajectories and on the union of MD and generated trajectories, yielding the corresponding projection operators $\mathcal{P}_{\mathtt{gen}}$ and $\mathcal{P}_{\mathtt{joint}}$, respectively.  For \textbf{MD PCA $\mathcal{W}_2$}, we apply the MD-based projector $\mathcal{P}_{\mathtt{md}}$ to the C$\alpha$ positions of both MD and generated trajectories, and compute the 2-Wasserstein distance between the resulting projected distributions. For \textbf{Joint PCA $\mathcal{W}_2$}, we apply the joint projector $\mathcal{P}_{\mathtt{joint}}$ to the C$\alpha$ positions of both MD and generated trajectories, and compute the 2-Wasserstein distance between the resulting projected distributions. The computation is defined as:
\begin{equation}
\begin{aligned}
\mathrm{MD\ PCA}\ \mathcal{W}_2
&=
\mathcal{W}_2
\bigl(
\mathcal{P}_{\mathtt{md}}(\mathcal{X}),
\mathcal{P}_{\mathtt{md}}(\mathcal{Y})
\bigr),
\\
\mathrm{Joint\ PCA}\ \mathcal{W}_2
&=
\mathcal{W}_2
\bigl(
\mathcal{P}_{\mathtt{joint}}(\mathcal{X}),
\mathcal{P}_{\mathtt{joint}}(\mathcal{Y})
\bigr).
\end{aligned}
\end{equation}
Besides, we compute the cosine similarity between the principal components obtained from MD $\mathcal{P}_{\mathtt{md}}$ and generated trajectories $\mathcal{P}_{\mathtt{gen}}$. Let $\mathbf{u}_1^{\mathtt{md}}$ and $\mathbf{u}_1^{\mathtt{gen}}$ denote the first principal components derived from PCA on the MD and generated trajectories, respectively. The cosine similarity is defined as:
\begin{equation}
\mathrm{PC\text{-}sim}=
\frac{
\left\langle
\mathbf{u}_1^{\mathtt{md}},
\mathbf{u}_1^{\mathtt{gen}}
\right\rangle
}{
\left \|
\mathbf{u}_1^{\mathtt{md}}
\right \|_2
\left \|
\mathbf{u}_1^{\mathtt{gen}}
\right \|_2
}.
\end{equation}
The PCA-based metric are reported by the median value across all test proteins.

\paragraph{Ensemble observables}
In addition, we evaluate the ensemble observables of our generated trajectories to assess whether our model captures the intermittent contacts and solvent exposure. We first identify the C$\alpha$ pairs and quantify the contact probability  with a 8 \AA threshold. Weak contacts are defined as those C$\alpha$ pairs that are in contact in the crystal structure(from the RCSB) but exhibit a dissociation probability  greater than 10\% in the trajectories. Transient contacts are defined as C$\alpha$ pairs that are not in contact in the crystal structure but associate with a contact probability exceeding 10\%. We further quantify the agreement between model generated and MD derived sets using the Jaccard similarity, named as Weak contacts $J$ and Transient contacts $J$. We compute the solvent-accessible surface area (SASA) of each side chain using the Shrake–Rupley algorithm with a probe radius of 2.8~\AA. A threshold of 2.0~\AA$^2$ is applied to distinguish buried from exposed residues. Exposed residues are defined as those whose side chains are buried in the crystal structure but become solvent-exposed in more than 10\% of structures in trajectories, and use Jaccard similarity to get exposed residue $J$.

\textbf{Weak / Transient contacts.}
Given a trajectory ensemble $\mathcal{X}=\{\mathbf{X}_m\}_{m=1}^{M}$, the contact probability for a residue pair $(n,k)$ is defined as
\begin{equation}
p_{(n,k)}(\mathcal{X})
=
\frac{1}{M}
\sum_{m=1}^{M}
\mathds{1}
\left(
\left\|
\mathbf{x}^{\mathrm{C}_\alpha}_{m,n}
-
\mathbf{x}^{\mathrm{C}_\alpha}_{m,k}
\right\|_2
\le 8~\text{\AA}
\right),
\qquad
1 \le n < k \le N .
\end{equation}

Given contact sets derived from generated trajectories $\mathcal{X}$ and MD trajectories $\mathcal{Y}$, weak and transient contacts are defined as
\begin{equation}
\begin{aligned}
\mathcal{C}_{\mathrm{weak}}^{\mathcal{X}}
&=
\left\{
(n,k)\in\mathcal{C}_{\mathtt{ref}}
\;\middle|\;
p_{(n,k)}(\mathcal{X}) < 0.9
\right\}, \\
\mathcal{C}_{\mathrm{trans}}^{\mathcal{X}}
&=
\left\{
(n,k)\notin\mathcal{C}_{\mathtt{ref}}
\;\middle|\;
p_{(n,k)}(\mathcal{X}) > 0.1
\right\}.
\end{aligned}
\end{equation}
where $\mathcal{C}_{\mathtt{ref}}$ denotes the set of contacting C$_\alpha$ residue pairs in the reference structure. We quantify their agreement using the Jaccard similarity and derive the \textbf{Weak / Transient contacts $J$} as
\begin{equation}
\begin{aligned}
J_{\mathrm{weak}}
&=
\frac{
\left|
\mathcal{C}_{\mathrm{weak}}^{\mathcal{X}}
\cap
\mathcal{C}_{\mathrm{weak}}^{\mathcal{Y}}
\right|
}{
\left|
\mathcal{C}_{\mathrm{weak}}^{\mathcal{X}}
\cup
\mathcal{C}_{\mathrm{weak}}^{\mathcal{Y}}
\right|
}, \qquad
J_{\mathrm{trans}}
&=
\frac{
\left|
\mathcal{C}_{\mathrm{trans}}^{\mathcal{X}}
\cap
\mathcal{C}_{\mathrm{trans}}^{\mathcal{Y}}
\right|
}{
\left|
\mathcal{C}_{\mathrm{trans}}^{\mathcal{X}}
\cup
\mathcal{C}_{\mathrm{trans}}^{\mathcal{Y}}
\right|
}.
\end{aligned}
\end{equation}
We report the median values across test proteins.

\textbf{Exposed residues.}
Let $\mathrm{SASA}_{m,n}$ denote the solvent-accessible surface area (SASA) of the side chain of residue $n$ at frame $m$, computed using the Shrake--Rupley algorithm with a probe radius of $2.8~\text{\AA}$. A residue is considered buried at frame $m$ if
\begin{equation}
\mathrm{SASA}_{m,n} < 2.0~\text{\AA}^2 .
\end{equation}

Let $\mathcal{B}_{\mathtt{ref}}$ denote the set of residues whose side chains are buried in the reference structure. The exposed residue set derived from a trajectory ensemble $\mathcal{X}=\{\mathbf{X}_m\}_{m=1}^{M}$ is defined as
\begin{equation}
\mathcal{E}^{\mathcal{X}}
=
\left\{
n \in \mathcal{B}_{\mathtt{ref}}
\;\middle|\;
\frac{1}{M}
\sum_{m=1}^{M}
\mathds{1}
\bigl(
\mathrm{SASA}_{m,n} > 2.0~\text{\AA}^2
\bigr)
> 0.1
\right\}.
\end{equation}

The agreement between exposed residue sets derived from generated trajectories $\mathcal{X}$ and MD trajectories $\mathcal{Y}$ is quantified using the Jaccard similarity, and the median values are reported:
\begin{equation}
J_{\mathrm{exp}}
=
\frac{
\left|
\mathcal{E}^{\mathcal{X}}
\cap
\mathcal{E}^{\mathcal{Y}}
\right|
}{
\left|
\mathcal{E}^{\mathcal{X}}
\cup
\mathcal{E}^{\mathcal{Y}}
\right|
}.
\end{equation}

\subsubsection{Structural Quality}
We adopt the precision and recall to examine the structure coverage with lDDT$_{\mathrm{C}\alpha}$: 
$\mathrm{lDDT}_{\mathrm{C}\alpha}=\frac{1}{N} \sum_{n=1}^{N} \frac{1}{4} \sum_{\delta \in [0.5, 1, 2, 4]}^{} \mathds{1}\left( \| x_{m,n} - y_{m, n} \| \leq \delta \right))$ where $N$ is the number of C$\alpha$ atoms in each structure $x_{m,n}$ or $y_{m,n}$
Specifically, the \textbf{precision} is the average lDDT$_{\mathrm{C}\alpha}$ . \textbf{Recall} is defined analogously, as the the average lDDT$_{\mathrm{C}\alpha}$ from each MD conformation $x_i$ to its most similar sampled conformation $y_{i}$. 
\textbf{Diversity} is the average dissimilarity (1-lDDT$_{\mathrm{C}\alpha}$) between any pairs of generated structure.

\paragraph{Structural Coverage}

Let $\{x_m\}_{m=1}^{M}$ be the set of generated conformations and $\{y_j\}_{m'=1}^{M'}$ be the set of MD conformations.
\textbf{Precision} is defined as the average lDDT$_{\mathrm{C}\alpha}$ from each generated conformation to its most similar MD conformation:
\begin{equation}
\mathbf{Precision}
=
\frac{1}{M}
\sum_{m=1}^{M}
\max_{1\le m'\le M'}
\mathrm{lDDT}_{\mathrm{C}\alpha}(x_m,y_{m'}).
\end{equation}
\textbf{Recall} is defined analogously as the average lDDT$_{\mathrm{C}\alpha}$ from each MD conformation to its most similar generated conformation:
\begin{equation}
\mathbf{Recall}
=
\frac{1}{M'}
\sum_{m'=1}^{M'}
\max_{1\le m\le M}
\mathrm{lDDT}_{\mathrm{C}\alpha}(x_m,y_{m'}).
\end{equation}

\textbf{Diversity} is quantified as the average dissimilarity between pairs of generated conformations:
\begin{equation}
\mathbf{Diversity}
=
\frac{2}{M\times (M-1)}
\sum_{1\le m<m'\le M}
\left(1-\mathrm{lDDT}_{\mathrm{C}\alpha}(p_m,p_{m'})\right).
\end{equation}

\paragraph{Stereochemical validity}
To assess stereochemical validity, we evaluate the C$\alpha$–C$\alpha$ and C–N distances between sequence-adjacent residues, using thresholds of 4.5~\AA\ and 2.0~\AA, respectively. We further compute distances between all pairs of backbone atoms belonging to different residues and define a steric clash as any interatomic distance below 1.0~\AA. For all selected distance-based metrics, values are averaged across relevant atom pairs and summarized using the median over the temporal dimension.

To assess stereochemical validity, we compute distance-based metrics on generated trajectories.
Let $\mathbf{x}_{m,n}^{\mathrm{C}_\alpha}$ denote the C$_\alpha$ coordinate of residue $n$ at frame $m$, and let $\mathbf{x}_{m,n}^{\mathrm{C}}$ and $\mathbf{x}_{m,n}^{\mathrm{N}}$ denote the backbone C and N atom coordinates, respectively.

For sequence-adjacent residues $(n,n+1)$, the C$_\alpha$–C$_\alpha$ and C–N distances are defined as
\begin{equation}
d^{\mathrm{C}_\alpha\text{-}\mathrm{C}_\alpha}_{m,n}
=
\left\|
\mathbf{x}_{m,n}^{\mathrm{C}_\alpha}
-
\mathbf{x}_{m,n+1}^{\mathrm{C}_\alpha}
\right\|_2,
\qquad
d^{\mathrm{C}\text{-}\mathrm{N}}_{m,n}
=
\left\|
\mathbf{x}_{m,n}^{\mathrm{C}}
-
\mathbf{x}_{m,n+1}^{\mathrm{N}}
\right\|_2 .
\end{equation}

Violations are identified by comparing these distances against thresholds of $4.5~\text{\AA}$ for C$_\alpha$–C$_\alpha$ distances and $2.0~\text{\AA}$ for C–N distances. A steric clash between two backbone atoms $a$ and $b$ belonging to different residues is defined as any interatomic distance satisfying
\begin{equation}
d^{\mathrm{clash}}_{m,(a,b)}
=
\left\|
\mathbf{x}_{m,a}
-
\mathbf{x}_{m,b}
\right\|_2
<
1.0~\text{\AA}.
\end{equation}

For each distance-based criterion, we compute the fraction of atom or residue pairs that satisfy the corresponding geometric constraint at each frame, and report the mean of these fractions over the temporal dimension for \texttt{Val-} metrics in Tab.~\ref{tab:performance}; we  reported the mean fraction of violations in Fig.~\ref{fig:MD_emulation}.

\paragraph{Long-timescale estimation}
Following ~\cite{bioemu2025}, we evaluate long-timescale performance by the mean absolute error (MAE) of protein free energy landscapes. Specifically, We apply time-lagged independent component analysis~\cite{perez2013identification}
to project the trajectories into 2D space with pairwise C$\alpha$ coordinates. The MAE between model generations and MD results is  then computed as:
\begin{equation}
\mathrm{MAE}=\frac{1}{N}\sum^N_i \mathrm{abs}(G^{\mathrm{gen}}_i-G^{\mathrm{MD}}_i)
\end{equation}
where $N$ denotes the number of generated trajectories. The free energy  $G^{\mathrm{gen}}$ and $G^{\mathrm{MD}}$ is estimated as:
\begin{equation}
G=-k_BT\ln (p_i)+\mathrm{const}
\end{equation}
where, $k_B$ is the Boltzmann constant, T is the temperature of the MD simulation, and $p_i$ is the normalized histogram probability.

\backmatter


\bibliography{sn-bibliography}

\end{document}